\documentclass[twocolumn, showpacs,prd]{revtex4}

\usepackage{epsfig}
\usepackage{graphicx}
\usepackage{dcolumn}
\usepackage{amsmath}
\usepackage{latexsym}

\begin{document}

%%%%%%%%%%%%%%%%%%%%
\title{Commutative/Non-Commutative Dualities}  
%\date{\today}
\author{FG Scholtz$^{a,b}$, PH Williams$^b$ and JN Kriel$^b$}
\affiliation{$^a$National Institute for Theoretical Physics (NITheP), 
Stellenbosch 7602, South Africa\\
$^b$Institute of Theoretical Physics, 
Stellenbosch University, Stellenbosch 7602, South Africa}

%%%%%%%%%%%%%%%%
\begin{abstract}
\noindent 
We show that it is in principle possible to construct dualities between commutative and non-commutative theories in a systematic way. This construction exploits a generalization of the exact renormalization group equation (ERG).  We apply this to the simple case of the Landau problem and then generalize it to the free and interacting non-canonical scalar field theory. This constructive approach offers the advantage of tracking the implementation of the Lorentz symmetry in the non-commutative dual theory.  In principle, it allows for the construction of completely consistent non-commutative and non-local theories where the Lorentz symmetry and unitarity are still respected, but may be implemented in a highly non-trivial and non-local manner.  
%%%%%%%%%%%%%%%%

\end{abstract}

\maketitle

%%%%%%%%%%%%%%%%%%%%%%%%%%%%%%%%%%%

%%%%%%%%%%%%%%%%%%%%%%%%%%%%%%%%%%%%%%%%%%%%%%%%%%%%%%%%%%%%

\section{Introduction}
\label{intro}

The structure of space-time at short length scales and the emergence of space-time as we perceive it at long length scales are probably the most challenging problems facing modern physics \cite{seib}.  These issues are also at the core of the struggle to combine gravity and quantum mechanics into a unified theory.  

One scenario for space-time at short length scales that has received considerable attention in the past few decades is that of non-commutative space-time.  This was originally proposed by Snyder \cite{Snyder} in an attempt to avoid the ultra-violet infinities of field theories.  The discovery of renormalization pushed these ideas to the background until more recently when they resurfaced in the search for a consistent theory of quantum gravity.  The compelling arguments of Doplicher et al \cite{dop} highlighted the need for a revised notion of space-time at short length scales and gave strong arguments in favour of a non-commutative geometry.  Shortly thereafter it was also noted that non-commutative coordinates occurred quite naturally in certain string theories \cite{wit}, generally perceived to be the best candidate for a theory of quantum gravity.  This sparked renewed interest in non-commutative space-time and the formulation of quantum mechanics \cite{scholtz} and quantum field theories on such spaces \cite{doug}.  

In these formulations the simplest form 
\begin{equation}
\label{ncom}
\left[\hat{x}_\mu , \hat{x}_\nu \right] = i \theta_{\mu \nu}
\end{equation}
for the space-time commutation relations is normally adopted, where $\theta_{\mu \nu}$ are constants.  Generalizations to fuzzy- and Snyder-space have been explored in \cite{pres,scholtz1,scholtz2,scholtz3} and \cite{bat} respectively.  The latter offers the possibility of avoiding the breaking of rotational and Lorentz symmetries.

In \cite{car1,car2} a slightly different proposal, often referred to as non-canonical quantum field theory, was made.  In this case non-commutativity is implemented on the level of the canonical commutation relations of the fields and momenta, rather than the space-time coordinates as is conventionally done \cite{doug}.  This is, indeed, also closer in spirit to non-commutative quantum mechanics as formulated in \cite{scholtz}.  

For both the constant space-time commutation relations of (\ref{ncom}) and in non-canonical quantum field theory the Lorentz symmetry gets violated. In the former case the Lorentz symmetry can be restored upon twisting \cite{chaichian}.  Despite this insight, several outstanding and controversial issues continue to plague the twisted implementation of the Lorentz symmetry \cite{pinzul}.  The first difficulty is to carry out the standard Noether analysis and identify conserved charges for the twisted Lorentz symmetry \cite{pinzul}.   Another obstacle is the quantisation of these theories, where one can either adopt the standard quantisation procedure or also deform the canonical commutation relations (see \cite{pinzul} and references therein).  On the level of the functional integral this amounts to altering the measure.  This has rather drastic consequences, such as the absence of UV/IR mixing.  Indeed, in  \cite{pinzul} it is argued that UV/IR mixing may be related to a quantum anomaly of the Lorentz group, which is very closely related to the choice of functional integral measure.  At this point there seems to be no consensus between these different points of view.

In the case of non-canonical quantum field theory a partial mapping to a commutative theory can restore the Lorentz invariance. However, this introduces non-local interactions for interacting theories and it is an open question whether these theories are unitary and renormalisable \cite{car3}.

In \cite{scholtz4} it was argued that the exact renormalization group (ERG) can be used to construct dualities between commutative and non-commutative quantum systems. This construction was carried out explicitly for a number of cases, particularly the Landau problem.  If this can be generalized to quantum field theories, the advantage is obvious: Although, as already pointed out above, the resulting non-commutative quantum field theories may be non-local and not obviously renormalisable, unitary or Lorentz invariant, the duality must ensure that these properties are still present, albeit in a non-manifest way.  This presents the possibility of constructing a wide class of quantum field theories, describing exactly the same physics as some standard quantum field theory, but in terms of non-commuting degrees of freedom.  The benefit that such dualities may offer are well known, particularly if they can be used to weaken interactions.  Indeed, that this is possible in the quantum Hall system has already been argued in \cite{scholtz5} and is further supported by the composite fermion picture.  An even greater advantage is to be gained if the duality construction can be carried out in a systematic way using the ERG, as was already demonstrated in \cite{scholtz4}.

This provides the motivation for the present paper, which aims to demonstrate and construct commutative/non-commutative dualities for field theories using the ERG program.  The focus here is on dualities between commutative and non-canonical quantum field theories as described in \cite{car1,car2,car3}.  Indeed, it turns out that these dualities are straightforward generalizations of the ones constructed in \cite{scholtz4}.  The paper is therefore organised as follows:  Section \ref{landau} explains the basic ideas, reviews and expands on the duality construction of \cite{scholtz4} for the Landau problem.  Section \ref{scalar} generalizes this construction to the case of a free complex scalar field theory and makes contact with the work done in \cite{car1,car2,car3}.  Section \ref{sym} discusses the fate of the Lorentz symmetry under the duality map and finally section \ref{int} generalizes the construction to the case of an interacting complex scalar field theory.

\section{Commutative/non-commutative dualities for the Landau problem}
\label{landau}

In this section we introduce and review the concepts we develop in this paper in the simplest possible case of the Landau problem.   The following sections generalize these ideas to the field theoretic setting.

Our concern for the rest of the paper will be the normalized generating functional from which all the correlation functions of the given theory can be computed.  In the case of a complex scalar field, which is the generic case considered here, this is given by 
\begin{equation}
\label{genfun}
Z[J]=\frac{W[J]}{W[0]}
\end{equation}
where
\begin{equation}
W[J]=\int [d{\bar\phi}][d\phi] e^{i\left[S[{\bar\phi},\phi]+{\bar J}\phi+J{\bar\phi}\right]}.
\end{equation}
Here ${\bar\phi}$ denotes complex conjugation. The action $S[{\bar\phi},\phi]$ is a real functional of the fields and we work in natural units with $\hbar=c=1$.  

For a particle of charge $-e<0$ moving in a magnetic field $B>0$ in the positive $z$-direction, and using complex coordinates $z=\frac{1}{\sqrt{2}}(x+iy)$, the action in the symmetric gauge is given by 
\begin{eqnarray}
\label{commact}
S=\int_{-\infty}^{\infty}dt &&\!\!\!\!\!\! \left[m\left\{\dot{\bar{z}}+\frac{ieB}{2m}\bar{z}\right\}
\left\{\dot{z}-\frac{ieB}{2m}z\right\}\right. \nonumber\\
&& \left. -\frac{e^2 B^2 }{4m}\bar{z}z\right].
\end{eqnarray} 

The corresponding action for the non-commutative Landau problem was derived in \cite{scholtz6,scholtz7}
\begin{eqnarray}
\label{ncact}
S=\int_{-\infty}^{\infty}dt &&\!\!\!\!\!\! \left[\frac{1}{2}\left\{\dot{\bar{z}}+\frac{ieB}{2m}\bar{z}\right\}\left(\frac{1}{2m}+\frac{i\theta}{2}
\partial_{t}\right)^{-1}
\left\{\dot{z}-\frac{ieB}{2m}z\right\}\right. \nonumber\\
&& \left. -\frac{e^2 B^2 }{4m}\bar{z}z\right].
\end{eqnarray} 
Here $\theta$ is the non-commutative parameter with the dimension of area.

Before proceeding, we consider the quantization of these theories.  For this it is convenient to return to the phase-space functional integral for which the action in the commutative case is given by
\begin{eqnarray}
\label{commphase}
S=\int_{-\infty}^{\infty}dt  &&\!\!\!\!\!\!  \left[\bar\pi\left(\dot{z}-\frac{ieB}{2m}z\right)+\left(\dot{\bar{z}}+\frac{ieB}{2m}\bar{z}\right)\pi\right. \nonumber\\
&& \left. -\frac{1}{m}\bar\pi\pi-\frac{e^2 B^2 }{4m}\bar{z}z\right]
\end{eqnarray}
and in the non-commutative case by
\begin{eqnarray}
\label{ncphase}
S=\int_{-\infty}^{\infty}dt &&\!\!\!\!\!\!\!\!\!\  \left[\bar\pi\left(\dot{z}-\frac{ieB}{2m}z\right)+\left(\dot{\bar{z}}+\frac{ieB}{2m}\bar{z}\right)\pi\right. \nonumber\\
&& \left. -\bar\pi\left(\frac{1}{m}+i\theta\partial_{t}\right)\pi-\frac{e^2 B^2 }{4m}\bar{z}z\right].%\nonumber \\
\end{eqnarray}
The functional integral is now over $\pi, \bar\pi$ and $z,\bar z$.  

The simplest way to carry out the quantization is to follow \cite{jackiw}.  For the commutative case this yields the commutation relations 
\begin{equation}
\label{comcom}
\left[z,\bar{\pi}\right]=\left[\bar{z},\pi\right]=i ,
\end{equation}
with all other commutators vanishing.  Clearly then the momenta and coordinates are individually commuting.  Using these commutation relations the Hamiltonian is found to be
\begin{equation}
\label{ham}
H=\frac{1}{m}\bar\pi\pi+\frac{e^2 B^2 }{4m}\bar{z}z+\frac{ieB}{2m}\left(\bar{\pi}z-\bar{z}\pi\right)-\frac{eB}{2m}
\end{equation}
where some care with ordering is required to obtain the correct form for the Zeeman term.  We can factorise $H$ as
\begin{equation}
\label{facham}
H=\frac{1}{m}\left(\bar\pi-\frac{ieB}{2}\bar{z}\right)\left(\pi+\frac{ieB}{2}z\right)-\frac{eB}{2m}.
\end{equation}
Introducing the boson creation and annihilation operators
\begin{equation}
a=\frac{1}{\sqrt{eB}}\left(\bar\pi-\frac{ieB}{2}\bar{z}\right),\quad a^\dagger=\frac{1}{\sqrt{eB}}\left(\pi+\frac{ieB}{2}z\right)
\end{equation}
yields 
\begin{equation}
H_C=\omega_c\left(a^\dagger a+\frac{1}{2}\right)
\end{equation}
where $\omega_c=\frac{eB}{m}$ is the cyclotron frequency.  As usual there is an infinite degeneracy in each Landau level, which stems from another set of creation and annihilation operators, constructed in terms of the guiding centre coordinates, which commute with $H$.

Following the same procedure for the non-commutative case, the corresponding commutation relations are
\begin{equation}
\label{nccom}
\left[z,\bar{\pi}\right]=\left[\bar{z},\pi\right]=i,\quad [z,\bar{z}]=\theta,
\end{equation}
with all other commutators vanishing.  Now the momenta are still commuting, but the coordinates are non-commuting, as one would have expected. The Hamiltonian is still given by (\ref{ham}) and the same factorization as in (\ref{facham}) applies.  Now, however, the creation and annihilation operators are 
\begin{equation}
a=\frac{1}{\sqrt{e|B^*|}}\left(\bar\pi-\frac{ieB}{2}\bar{z}\right),\quad a^\dagger=\frac{1}{\sqrt{e|B^*|}}\left(\pi+\frac{ieB}{2}z\right)
\end{equation}
when the effective magnetic field 
\begin{equation}
\label{effB}
B^*=B\left(1-\frac{eB\theta}{4}\right)
\end{equation}
is positive. For negative $B^*$ the definitions of $a$ and $a^\dagger$ are exchanged. The Hamiltonian becomes 
\begin{equation}
\label{ncham}
H_{NC}=\omega^*_c a^\dagger a+\omega^*_c-\frac{\omega_c}{2}, 
\end{equation}
where $\omega^*_c$ is the effective cyclotron frequency $\omega^*_c=\frac{e|B^*|}{m}$.  There is again an infinite degeneracy of each Landau level.  

We now turn to the duality between the commutative and non-commutative theories, i.e we want to establish a direct relationship between the generating functional (\ref{genfun}) for the commutative action (\ref{commact}) and the non-commutative action (\ref{ncact}).  There are two ways in which this can be done. The first makes use of a non-canonical change of variables in the functional integral and the second uses the exact renormalization (ERG) to establish this relation.  Although the former is more transparent and physically intuitive, the latter allows for a more general constructive approach to these dualities, which can also be applied to interacting systems. We therefore demonstrate both approaches here for benchmarking and maximal clarity.

Let us start with the non-canonical transformation.  We derive a slightly more general duality then just a commutative/non-commutative one.  We introduce the change of variables 
\begin{equation}
\label{ncchange}
z=Z-ia\pi,\quad \bar{z}=\bar{Z}+ia\bar{\pi}
\end{equation}
in the \emph{non-commutative} phase-space generating functional integral with action (\ref{ncphase}). This yields 
\begin{eqnarray}
\label{nctrans}
S=\int_{-\infty}^{\infty}dt &&\!\!\!\!\!\!\!\!\!\  \left[\bar\pi\left(\dot{Z}-\frac{ie\tilde{B}}{2\tilde{m}}Z\right)+\left(\dot{\bar{Z}}+\frac{ie\tilde{B}}{2\tilde{m}}\bar{Z}\right)\pi\right. \nonumber\\
&& \left. -\bar\pi\left(\frac{1}{\tilde{m}}+i\tilde{\theta}\partial_{t}\right)\pi-\frac{e^2 \tilde{B}^2 }{4\tilde{m}}\bar{Z}Z\right.\nonumber \\
&&\left. +\bar{J}\left(Z-ia\pi\right)+\left(\bar{Z}+ia\bar{\pi}\right)J\right]
\end{eqnarray}
where
\begin{equation}
\label{renp}
\tilde{B}=\frac{2B}{2+aeB},\quad \tilde{m}=\frac{4m}{(2+aeB)^2},\quad \tilde{\theta}=\theta+2a.
\end{equation}
Upon comparison with (\ref{ncphase}) we immediately recognize this as another non-commutative system with $\tilde{\theta}=\theta+2a$.  The only difference from the standard non-commutative generating functional is the coupling between $\pi$ and the source $J$.

We now attempt to eliminate these unwanted couplings.  To do this we introduce a further change of variables by setting
\begin{equation}
%\tilde{\pi}=\pi-ia\Delta^{-1}J,
\pi=\tilde{\pi}+ia\Delta^{-1}J,
\end{equation}
where we have simplified notation by setting
\begin{equation}
\Delta=\left(\frac{1}{\tilde{m}}+i\tilde{\theta}\partial_{t}\right).
\end{equation}
Following this change of variables we find that the $\pi$-$J$ couplings can be eliminated in favour of standard renormalized source terms and a quadratic source term, provided we set
\begin{equation}
\label{a}
 a=\frac{1}{e\tilde{B}}-\frac{\theta}{2}.
 \end{equation}
This yields
\begin{eqnarray}
\label{nctrans1}
S=\int_{-\infty}^{\infty}dt &&\!\!\!\!\!\!\!\!\!\  \left[\bar{\tilde{\pi}}\left(\dot{Z}-\frac{ie\tilde{B}}{2\tilde{m}}Z\right)+\left(\dot{\bar{Z}}+\frac{ie\tilde{B}}{2\tilde{m}}\bar{Z}\right)\tilde{\pi}\right. \nonumber\\
&& \left. -\bar{\tilde{\pi}}\left(\frac{1}{\tilde{m}}+i\tilde\theta\partial_{t}\right)\tilde{\pi}-\frac{e^2 \tilde{B}^2 }{4\tilde{m}}\bar{Z}Z\right.\nonumber \\
&&\left. +\left(1-\frac{a}{\tilde{\theta}}\right)\bar{J}Z+\left(1-\frac{a}{\tilde{\theta}}\right)\bar{Z}J\right.\nonumber\\
&&\left. +a^2\bar{J}\Delta^{-1}J\right].
\end{eqnarray}

Equations (\ref{renp}) and (\ref{a}) can be solved to yield
%\begin{eqnarray}
%\tilde{\theta}&&=\frac{4-\theta eB}{eB},\nonumber\\
%\tilde{B}&&=\frac{2B}{4-\theta eB},\nonumber\\
%\tilde{m}&&=\frac{4m}{(4-\theta eB)^2},\nonumber\\
%a&&=\frac{2-\theta eB}{eB}.
%\end{eqnarray}
\begin{align}
\tilde{\theta}&=\frac{4-\theta eB}{eB} & \tilde{B}&=\frac{2B}{4-\theta eB}\\
\tilde{m}&=\frac{4m}{(4-\theta eB)^2} & a&=\frac{2-\theta eB}{eB}
\end{align}

We now have the exact relation between the generating functionals of two non-commutative theories
\begin{equation}
Z_\theta\left[J\right]=e^{ia^2\int_{-\infty}^{\infty}dt  \left[\bar{J}\Delta^{-1}J\right]}Z_{\tilde{\theta}}\left[\kappa J\right]
\end{equation}
with
\begin{equation}
\kappa=\frac{2}{4-\theta eB}.
\end{equation}

In the special case of $\theta=0$, this gives the commutative/non-commutative duality
\begin{equation}
\label{dual}
Z_c\left[J\right]=e^{ia^2\int_{-\infty}^{\infty}dt  \left[\bar{J}\Delta^{-1}J\right]}Z_{nc}\left[\frac{1}{2} J\right]
\end{equation}
where the non-commutative theory has non-commutative parameter $\tilde{\theta}=\frac{4}{eB}$.

This establishes a complete duality between the commutative and non-commutative theories.  Note that no zero mass limit or infinite magnetic field limit had to be taken as is normally done, but that this is an exact duality.  

From (\ref{dual}) we note the following simple rule for the computation of correlators in the commutative theory in terms of correlators of the non-commutative theory
\begin{eqnarray}
\frac{\delta}{\delta J}&&\rightarrow \frac{\delta}{\delta J}+ia^2\bar{\Delta}^{-1}\bar{J},\nonumber\\
\frac{\delta}{\delta \bar{J}}&&\rightarrow \frac{\delta}{\delta \bar{J}}+ia^2\Delta^{-1} J.
\end{eqnarray}
This can also be easily inverted to find the correlators of the non-commutative theory in terms of the commutative theory.  We now have a complete dictionary for the computation of correlation functions.

The commutative/non-commutative duality that transpired above is in fact much more general as has been anticipated in \cite{scholtz8} and has been discussed in detail for the Landau problem in \cite{scholtz4}. The core idea in the constructive approach of \cite{scholtz4} is to use the ERG to construct dual families of non-commutative theories.  The key observation is that the commutative and non-commutative theories in essence differ in the kinetic energy term, as is clear from comparing (\ref{commact}) with (\ref{ncact}).  One can thus take the point of view, as is done in the ERG approach, that the kinetic energy term of the commutative theory is modified in conjunction with the interacting part such that the normalized generating functional is invariant.  The only difference from the standard ERG approach is that the restricting assumptions on the modified kinetic energy term and source terms do not necessarily apply in this case.  This can be addressed easily by a complimentary set of flow equations for the sources term, derived in \cite{scholtz4}.  The details of this construction for the Landau problem is discussed in \cite{scholtz4}, and so we only outline the basic steps here.

We consider a complex scalar field theory in $0+1$-dimensions with Fourier transformed action
\begin{eqnarray}
\label{action}
S[\phi, \bar\phi]=\int d\omega\,\bar\phi(\omega)K(\omega, \ell)\phi(\omega)+S_{I}[\phi, \bar\phi]+J_{\ell}[\phi, \bar\phi].\nonumber\\
\end{eqnarray}
Here $K(\omega, \ell)$ takes the standard form $m\omega^2$ at $\ell=0$ and $J_{\ell}[\phi, \bar\phi]$ is a generalised source term, which is a functional of the fields, determined by the requirement of invariance of the generating functional. It will be sufficient to take this source term to be linear:
\begin{eqnarray}
J_{\ell}[\phi, \bar\phi]=\int d\omega~[J_{0}(l) +\bar{J}_{0}(l) +J_{1}(l)\bar\phi(\omega)+\bar{J}_{1}(l)\phi(\omega)].\nonumber\\.
\label{2}
\end{eqnarray}
Here $J_{0}(0)=\bar{J}_{0}(0)=0$ and 
%The initial conditions on the source term are
%\begin{eqnarray}
%\label{2x}
%J_{0}(0)&=&\bar{J}_{0}(0)=0, \\
%J_{1}(\ell)|_{\ell=0}=J_{1}(0)&,&\bar{J}_{1}(\ell)|_{\ell=0}=\bar{J}_{1}(0).
%\label{2y}
%\end{eqnarray}
$J_1(0)$ is an arbitrary function of $\omega$ that acts as the source in the bare ($\ell=0$) action.   In what follows we denote the first term in eq.(\ref{action}) by $S_{0}[\phi, \bar\phi]$ and refer to the second term $S_{I}[\phi, \bar\phi]$ as the interacting part of the action. Both are functions of the flow parameter $\ell$.

We now apply the logic of the ERG as set out in \cite{pol} and \cite{banks}, and require $Z[J]=W[J]/W[0]$ to be invariant under the flow, i.e. independent of $\ell$.  However, we relax the usual conditions imposed on $K(\omega, \ell)$ and the sources, and this necessitates the flow of the source terms as well.  A derivation similar to the one carried out in \cite{banks} yields the following equations for the interacting part and source terms:
\begin{eqnarray}
\label{9}
\partial_{\ell}S_{I}=\int d\omega~\partial_{\ell}K^{-1}
\left\{\frac{\delta S_{I}}{\delta \bar\phi(\omega)}\frac{\delta S_{I}}{\delta \phi(\omega)}
-\frac{\delta^{2} S_{I}}{\delta \bar\phi(\omega)\delta \phi(\omega)}\right\}\nonumber\\
\end{eqnarray}
\begin{eqnarray}
\partial_{\ell}J_{\ell}&=&\int d\omega~\partial_{\ell}K^{-1}
\left\{\frac{\delta S_{I}}{\delta \phi(\omega)}\frac{\delta J_{\ell}}{\delta \bar\phi(\omega)}
+\frac{\delta S_{I}}{\delta \bar\phi(\omega)}\frac{\delta J_{\ell}}{\delta \phi(\omega)}\right.\nonumber\\
&&\left.+\frac{\delta J_{\ell}}{\delta \bar\phi(\omega)}\frac{\delta J_{\ell}}{\delta \phi(\omega)}
-\frac{\delta^{2} J_{\ell}}{\delta \bar\phi(\omega)\delta \phi(\omega)}\right\}.
\label{10}
\end{eqnarray}

These equations can easily be solved when the interaction term is quadratic in the fields, i.e. 
\begin{eqnarray}
S_{I}[\phi, \bar\phi]=\int d\omega\,\bar\phi(\omega) g(\omega, \ell) \phi(\omega)
\label{11}
\end{eqnarray}
with $g(\omega, \ell)$ real.  
%In this case it is also easy to see that the generalized source terms are linear:
%\begin{eqnarray}
%J_{\ell}[\phi, \bar\phi]=\int d\omega~[J_{0}(l) +\bar{J}_{0}(l) +J_{1}(l)\bar\phi(\omega)+\bar{J}_{1}(l)\phi(\omega)].\nonumber\\.
%\label{2}
%\end{eqnarray}
Focussing on the source terms for the moment and using eqs.(\ref{10}) and (\ref{2}) yields
\begin{eqnarray}
\label{13a}
\partial_{\ell}J_{1}(\ell)&=&\partial_{\ell}K^{-1}(\omega, \ell)g(\omega, \ell)  J_{1}(\ell),\nonumber\\
\label{13b}
\partial_{\ell}[J_{0}(\ell)+\bar{J}_{0}(\ell)]&=&\partial_{\ell}K^{-1}(\omega, \ell)|J_{1}(\ell)|^{2}.
\label{13c}
\end{eqnarray}
Integrating these equations and using the initial conditions on the sources, we obtain
\begin{align}
\label{14a}
J_{1}(\ell)&=J_{1}(0)\exp\left(\int_{0}^{\ell} d\ell'~g(\omega, \ell')\partial_{\ell'}K^{-1}(\omega, \ell')\right),\nonumber\\
%\label{14b}
J_{0}(\ell)+\bar{J}_{0}(\ell)&=|J_{1}(0)|^{2}\int_{0}^{\ell} d\ell'~\partial_{\ell'}K^{-1}(\omega, \ell')\nonumber\\
&\times\exp\left(2\int_{0}^{\ell'} d\ell''~g(\omega, \ell'')\partial_{\ell''}K^{-1}(\omega, \ell'')\right).
%\label{14c}
\end{align}

We now apply this result to the Landau problem (\ref{commact}), for which the Fourier transformed action reads
\begin{eqnarray}
S&=&\int d\omega\,[\bar{z}(\omega)m\omega^2 z(\omega) -eB\omega\bar{z}(\omega)z(\omega) \nonumber\\
&&+J(\omega)\bar{z}(\omega)+\bar{J}(\omega)z(\omega)].
\label{15}
\end{eqnarray}

Motivated by the form of the action for the non-commutative Landau problem we take
\begin{eqnarray}
K(\omega, \ell)=\frac{m\omega^2}{(1-m\omega\ell)},
\label{17}
\end{eqnarray}
and the interacting part of the action as in (\ref{11}) with initial condition
\begin{eqnarray}
g(\omega, \ell)|_{\ell=0}=-eB\omega.
\label{18}
\end{eqnarray}

This yields the dual action \cite{scholtz4} 
\begin{eqnarray}
S&=&\int d\omega\left[\bar{\tilde z}(\omega)K(\omega, \ell) \tilde{z}(\omega)
 -\frac{eB\omega}{(1-m\omega\ell)}\bar{\tilde z}(\omega)\tilde z(\omega)\right.
\nonumber\\
&&\left.-\frac{|J_{1}(0)|^{2}\ell}{\omega(1-eB\ell)}
+\left(\frac{J_{1}(0)}{\sqrt{1-eB\ell}}\bar{\tilde{z}}(\omega)+c.c.\right)\right].\label{28}
\end{eqnarray}
where 
\begin{equation}
\tilde{z}(\omega)=\sqrt{\frac{2}{1-eB\ell}}z(\omega).
\end{equation}
We recognise this as the action of a non-commutative Landau problem with $\theta=\ell$ and a magnetic field $B^*$ determined by
\begin{eqnarray}
B=B^{*}\left(1-\frac{eB^{*}\ell}{4}\right).
\label{30}
\end{eqnarray}
%One therefore again obtains a duality of the form (\ref{dual}) and thus (\ref{28}) represents a family of dual non-commutative theories, generalizing the result obtained through the non-canonical transformation where the non-commutative parameter was fixed.  
One therefore again obtains a duality similar in form to (\ref{dual}), and thus (\ref{28}) represents a family of dual non-commutative theories. This duality should be interpreted as follows: For every commutative system with magnetic field $B$ there is a corresponding family of non-commutative systems with non-commutative parameter $\theta=\ell$ and magnetic field $B^*$. Using these values of the non-commutative parameter and magnetic field in the excitation energy  (\ref{effB}) indeed yields the commutative cyclotron frequency $\omega_c=\frac{eB}{m}$ as is required by the duality.   One can also verify by explicit computation that the generating functionals are indeed identical.  It should also be noted that the duality is not unique, but can parameterized in many different ways as explained in \cite{ros}.

\section{Commutative/non-commutative dualities for a free complex scalar field theory}
\label{scalar}
We now proceed to apply the ideas of the previous section to construct dualities between commutative and non-commutative complex scalar field theories, where the non-commutative theory is a non-canonical field theory as described in \cite{car1,car2,car3}.  

We start with the free non-canonical complex scalar field theory for which the fields satisfy the equal time commutation relations
\begin{eqnarray}
\label{ncscal}
&&\left[\hat{\phi}(x), \hat{\phi}^\dagger(y)\right] = \theta\delta(x-y)\nonumber\\
&&\left[\hat{\phi}(x), \hat{\Pi}^\dagger(y)\right] = \left[\hat{\phi}^\dagger(x), \hat{\Pi}(y)\right] =i\delta(x-y)
\end{eqnarray}
with all other commutators vanishing.  In \cite{car3} the corresponding action is derived and found to be
\begin{equation}
\label{ncactionsc}
S = \int d^4 x \; \bar\phi(t,\vec{x})\left( \frac{-\partial_t^2}{1 + i \theta \partial_t - i \epsilon} + \vec{\nabla}^2 - m^2\right) \phi(t,\vec{x}).
\end{equation}

As a consistency check, we quantize this theory to verify that the commutation relations (\ref{ncscal}) are indeed reproduced.  Introducing auxiliary fields $\chi$ and $\bar{\chi}$ we can rewrite the action as
\begin{eqnarray}
S = \int d^4 x \; &&\!\!\!\!\!\!\left[\bar\chi \left( 1 + i \theta \partial_t - i \epsilon \right) \chi + \bar{\chi} \partial_t \phi + \chi \partial_t \bar\phi \right.\nonumber \\
&& \left. + \bar\phi (\vec{\nabla}^2 - m^2) \phi\right].
\end{eqnarray}
The epsilon term ensures the convergence of the above integral, but hereafter we drop it. We follow the Faddeev-Jackiw method \cite{jackiw} to quantize the system.  We denote
\begin{equation}
\xi^i = \{\phi,\bar\phi,\chi,\bar{\chi} \}
\end{equation}
and write the Lagrangian as
\begin{equation}
\int d^3 x \; \left[ \frac{1}{2} \xi^i \omega_{i j} \partial_t \xi^j - V(\xi)\right]
\end{equation}
with
\begin{equation}
\omega_{i j} = \begin{pmatrix}
0 & 0 & 0 & -1  \\
0 & 0 & -1 & 0  \\
0 & 1 & 0 & i \theta  \\
1 & 0 & - i \theta & 0
\end{pmatrix}.
\end{equation}
Following \cite{jackiw} the commutation relations (\ref{ncscal}) are indeed recovered from the entries of $i\omega^{ij}$ with $\omega^{ij}$ the inverse of $\omega_{i j}$. The Hamiltonian has the form
\begin{equation}
H = V(\xi) = \int d^3 x \; \left[\hat{\chi}^\dagger \hat{\chi} - \hat{\phi}^\dagger (\vec{\nabla}^2 - m^2) \hat{\phi}\right].
\end{equation}
The auxiliary fields are constrained to the conjugate momenta, yielding the final Hamiltonian
\begin{equation}
H = \int d^3 x \; \hat{\Pi}^\dagger \hat{\Pi} + \vec{\nabla}\hat{\phi}^\dagger \vec{\nabla}\hat{\phi} + m^2 \hat{\phi}^\dagger \hat{\phi}
\end{equation}
which can be second quantized in the standard way \cite{car3}.

Our aim now is to construct a duality between the free commutative complex scalar theory and the non-canonical complex scalar theory above.  To do this we follow closely the ERG procedure described in section \ref{landau}.   As mentioned earlier, the key idea is to modify the kinetic energy term in conjunction with the interaction, while leaving the normalized generating functional unchanged.  In this case we want to map to the non-canonical complex scalar field theory, and so we modify the kinetic energy of the commutative theory to match that of the non-canonical theory.  In Fourier space (\ref{ncactionsc}) reads
\begin{equation}
%\label{ncactionsc}
S = \int d^4 k \; \bar\phi(k^0,\vec{k})\left( \frac{\left(k^0\right)^2}{1 - \theta k^0 } - \left( \vec{k}^2 + m^2\right)\right) \phi(k^0,\vec{k}),
\end{equation}
which amounts to the modification
\begin{equation}
(k^0)^2 \rightarrow K(\theta) = \frac{(k^0)^2}{1 - \theta k^0}.
\end{equation}
Solving the ERG as in \cite{scholtz8}, the remaining quadratic term must flow as 
\begin{equation}
-\left(\vec{k}^2 + m^2 \right) \rightarrow g(\theta) = \frac{k^0 \left(\vec{k}^2 + m^2\right)}{\theta(\vec{k}^2 + m^2) - k^0}.
\end{equation}
The source term becomes
\begin{equation}
J_1(0) \rightarrow J_1(\theta) = \frac{k^0 J_1(0)}{k^0 - \theta(\vec{k}^2 + m^2)} 
\end{equation}
with an additional quadratic term
\begin{equation}
J_0(\theta) = -\frac{\theta |J_1(0)|^2}{k^0 - \theta(\vec{k}^2 + m^2)}.
\end{equation}
The full action with source terms is therefore
\begin{equation}
S_\theta = \int d^4 k \; \left[\bar\phi K \phi + \bar\phi g \phi + J[\phi,\bar\phi]\right].
\end{equation} 

This action is not quite that of the non-commutative, non-canonical system. However, it can easily be brought into the desired form with an additional momentum dependent rescaling of the fields:
\begin{equation}
\label{rescale}
\phi(k^0,\vec k) \rightarrow\sqrt{\frac{K(\theta)+g(0)}{K(\theta)+g(\theta)}}\phi(k^0,\vec k).
\end{equation}
This of course also modifies the sources.  Now we have again a complete duality between the commutative and non-canonical field theories in the sense of a precise relation between generating functionals just like in section \ref{landau}.

As explained in \cite{scholtz4} and \cite{ros} this result can also be interpreted as a blocking procedure. To see this, recall that it is possible to interpret the renormalization flow as a change of variables in the path integral.  Bar the source terms this is in our case a rather simple change of variables involving a momentum dependent scaling of the commutative fields:
\begin{equation}
\label{coarse}
\phi_0 (k^0,\vec{k})=\sqrt{\frac{K(\theta) + g(\theta)}{K(0) + g(0)}}\phi_{\theta}(k^0,\vec{k}).
\end{equation}
This immediately yields the non-commutative action with modified sources, which encode the dictionary between the two descriptions. Note that in position space the momentum dependent rescaling above is highly non-local and corresponds to a coarse graining as described in \cite{ros}.   In the interacting case, this becomes a much more complicated transformation.  

\section{Symmetries}
\label{sym}
Before we turn to the interacting case, let us briefly consider the fate of the Lorentz symmetry under the dualities constructed via the ERG and corresponding to a coarse graining as in section \ref{scalar}.  Clearly, if the two theories are truly dual, as is ensured by the ERG flow equations, the symmetry must be present in both, although it may not be manifest in the dual theory and implemented in a highly non-trivial way due to the coarse graining not being manifestly Lorentz covariant.  In fact, it should already be clear from (\ref{coarse}) that the Lorentz symmetry cannot be manifest in the dual theory and must be implemented in a highly non-local manner.  We now proceed to discuss this more generally in the setting of the simplest type of coarse graining that involves a momentum dependent rescaling of the fields, such as in (\ref{coarse}).

To do this, we recall a more general formulation of the flow equation which also establishes a direct link with coarse graining \cite{ros}: 
\begin{equation} \label{ren}
-\partial_\theta e^{- S_\theta[\phi]} = \int d^4 k \; \frac{\delta}{\delta \phi(k)} \left(\Psi_\theta(k) e^{- S_\theta[\phi]}\right) ,
\end{equation}
where $\Psi_\theta$ is a general functional of the fields.  If the action satisfies this flow equation, the normalized generating functional is again invariant if the sources are also adjusted appropriately.  

Every allowed $\Psi_\theta$ is related to some blocking transformation (coarse graining) \cite{ros}. If we write the blocking transformation as $f_\theta[\phi_0] = \phi$, where $\phi_0$ and $\phi$ are the commutative and non-commutative fields, respectively, then
\begin{equation}
\Psi_\theta(k) e^{- S_\theta[\phi]} = \int [d \phi_0] \; \delta\left[\phi - f_\theta[\phi_0]\right] \left( \partial_\theta f_\theta[\phi_0] \right) e^{-S_0}.
\end{equation}
For the simplest coarse graining which is just multiplicative and invertible $f_\theta[\phi_0](k) = f_\theta(k)\phi_0(k)$, we have $\Psi_\theta(k) = (\partial_\theta f_\theta(k))f^{-1}_\theta(k) \phi(k)$. Then equation \eqref{ren} gives a simple equation for the flow of the sources,
\begin{equation}
\partial_\theta J_\theta = (\partial_\theta f_\theta(k))f^{-1}_\theta(k) J_\theta.
\end{equation}

The symmetries are implemented differently in the two theories.  If the fields transform as $\phi_0 \rightarrow \tilde{\phi}_0$ under a symmetry transformation in the original theory, the transformation induced in the dual theory is
\begin{equation}
\phi \rightarrow \tilde{\phi} =  f_\theta(k)\tilde{\phi}_0(k).
\end{equation} 
Applying this to the Lorentz generators, we find them to be given in the dual theory by
\begin{equation}
M^{i j} = k^i \tilde{\partial}_{k^j} - k^j \tilde{\partial}_{k^i} 
\end{equation}
where
\begin{equation}
\tilde{\partial}_{k^i} =  \partial_{k^i} - \frac{\partial_{k^i} f(k)}{f(k)}.
\end{equation}
Note that in real space this corresponds to a highly non-local implementation of the Lorentz symmetry. 

\section{Commutative/non-commutative dualities for an interacting complex scalar field theory}
\label{int}

In this section we consider the construction of commutative/non-commutative dualities for a $\phi^4$ theory with interaction term
\begin{eqnarray}
\lambda  \int \left(\prod^4_{i = 1} d^4 k^i \right) \!\!\!\!\!\!\; &&\bar\phi(k^1) \phi(k^2) \bar\phi(k^3) \phi(k^4)\nonumber\\
&&\times \delta(k^1 - k^2 + k^3 - k^4).
\end{eqnarray}
Interactions of all possible orders in $\phi$ can now be generated during the flow and the individual coupling strengths and sources flow according to (\ref{9}) and (\ref{10}). We will simplify notation by setting
\begin{equation}
\partial_\theta K^{-1}(k) = F(k),
\end{equation}
\begin{widetext}
\begin{eqnarray}
&g_2(\theta,\lambda,k) = F(k) g_2^2(k) -  4\int d^4 p \;  F(p) g_4(k,p,k,p), \nonumber \\
&g_4(\theta,\lambda,k^1,k^2,k^3,k^4)= \sum_{i=1}^4 F(k^i) g_2 (k^i) g_4(k^1,k^2,k^3,k^4) -  9 \int d^4 p \; F(p) g_6(k^1,k^2,p,k^3,k^4,p)\\
&\vdots& \nonumber
\end{eqnarray}
\end{widetext}
It is not possible to solve these coupled differential equations simultaneously. We can, however, solve them perturbatively in the interaction strength. In orders of $\lambda$ we have $g_n = g_n^{(0)} + \lambda g_n^{(1)} + \ldots$ We can see all the interactions above $g_4$ are of order $\lambda^2$ or higher, and due to the initial conditions, they must vanish when $\theta$ goes to zero. We note that there is no zeroth order term for $g_4$. Solving the differential equations then produces:
\begin{eqnarray}
g_2^{(0)}(k) &=& \frac{k^0 \left(\vec{k}^2 + m^2\right)}{\theta(\vec{k}^2 + m^2) - k^0} \\
g_2^{(1)}(k) &=& \frac{-4 c \theta (k^0)^2}{(\theta(\vec{k}^2+m^2) - k^0)^2}  \\
g_4(k^1,k^2,k^3,k^4) &=& \prod_{i=}^4 \frac{(k^0)^i}{\theta((\vec{k}^i)^2 + m^2) - (k^0)^i}
\end{eqnarray}
where
\begin{equation}
c = \int d^4 p \; \frac{p^0}{(\vec{p}^2 + m^2 - p^0)^2}
\end{equation}
is a divergent integral that has to be regularized appropriately. 

Again the sources flow, but now the source terms are higher than linear order in the fields. To order $\lambda$ the sources have the form $J[\phi,\bar\phi] =J^{(0)}[\phi,\bar\phi]  + \lambda J^{(1)}[\phi,\bar\phi]$, where
\begin{equation}
J^{(0)}[\phi,\bar\phi] = \int d^4 k\left[J_0^{(0)}(k) + J_1^{(0)}(k) \bar\phi(k) + c.c. \right]
\end{equation}
as before and
\begin{eqnarray}
&& J^{(1)}[\phi,\bar\phi] = \int d^4 k \; \left[ J_0^{(1)}(k) + J_1^{(1)}(k) \bar\phi(k) + c.c. \right] + \nonumber \\  && \int d^4 k^1 d^4 k^2 \; \left[ J_2(k^1,k^2) \bar{\phi}(k^1) \bar{\phi}(k^2) + c.c.\right] + \nonumber \\ && \int d^4 k^1 d^4 k^2 \; \left[ \tilde{J}_2(k^1,k^2) \bar\phi(k^1) \phi(k^2) \right] + \nonumber \\ && \int d^4 k^1 d^4 k^2 d^4 k^3 \;  \left[J_3(k^1,k^2,k^3) \bar\phi(k^1) \bar\phi(k^2) \phi(k^3) +c.c.\right]\nonumber\\
\end{eqnarray}
where $\tilde{J}_2$ is real. The flow equations are included in appendix \ref{app}. In principle it is possible to solve all these equations, up to some divergences. For example:
\begin{eqnarray}
J_3(k^1,k^2,k^3) =&& \prod_{i=1}^3 \frac{(k^0)^i}{\theta((\vec{k}^i)^2 + m^2) - (k^0)^i}\times \nonumber\\ 
&&\ln \left( \frac{\kappa^0}{\theta(\vec{\kappa}^2 + m^2) - \kappa^0} \right) J_1^{(0)}(\kappa)\nonumber
\end{eqnarray}
where
\begin{equation}
\kappa = k^1 - k^2 - k^3.
\end{equation}

Finally, rescaling the fields as in (\ref{rescale}) transforms the quadratic part of the non-commutative action into the standard form (\ref{ncactionsc}) and thus establishes a complete duality between the interacting commutative and non-commutative theories, albeit in a perturbative way.  Note that although the interactions of the non-commutative theory are highly non-local the duality should ensure, at least in principle, that both the Lorentz symmetry and unitarity are maintained. To what extent these dualities can be useful requires further exploration.

\section{Conclusions}
\label{conc}

The aim of this paper was to demonstrate, at least in principle, how the ERG philosophy can be used to construct dualities between commutative and non-commutative theories.  The essence of the construction lies in noting that non-commutativity amounts to a modification of the kinetic energy term of the action.  From the ERG point of view, this change can be countered by an appropriate modification of the interacting part of the theory, captured by the ERG flow equation, such that the normalized generating functional remains invariant.  Here no limiting assumptions are made on the sources as is usually done in the ERG, and this requires the source terms to also flow.  

This construction was demonstrated in the simplest setting of the Landau problem and then generalized to free and interacting complex scalar field theories.  %The key advantages offered by this approach are (1) a constructive procedure, (2) a systematic way to tract the implementation of the Lorentz symmetry in the dual theory, which may not be manifestly Lorentz invariant, and (3) a systematic way of constructing interactions in the dual theory where the apparent lack of Lorentz symmetry makes it difficult to construct consistent interacting theories \cite{car3}.
The key advantages of this approach are that it provides a constructive procedure which allows us to trace the implementation of the Lorentz symmetry in the dual theory, which may not be manifestly Lorentz invariant. Furthermore, it allows for a systematic way of constructing interactions in the dual theory where the apparent lack of Lorentz symmetry makes it difficult to introduce interactions consistently \cite{car3}.

The mere existence of such dualities does not necessarily bring an obvious benefit.  Further exploration is required to determine whether this constructive approach may offer real advantages.  Such investigations may also have to move beyond the idea of purely commutative/non-commutative dualities.\\  

\noindent {\bf{Acknowledgements}}: This work was supported under a grant of the National Research Foundation of South Africa.

\appendix
\section{Flow of the Sources}
\label{app}

As before the flow for the zeroth order in $\lambda$ sources is given by 
\begin{eqnarray}
& J_0^{(0)}(k) = F(k) |J_1^{(0)}(k)|^2, \nonumber\\
&J_1^{(0)}(k) = F(k) g^{(0)}_2(k) J_1^{(0)}(k). \nonumber
\end{eqnarray}
For the higher order terms
\begin{widetext}
\begin{eqnarray}
&&J_0^{(1)}(k) = F(k) \left( \left( \bar{J}_1^{(1)}(k) J_1^{(0)}(k) + c.c. \right) +  \tilde{J}_2(k,k) \right)   ,\nonumber \\
&&J_1^{(1)}(k) =F(k) \left(g_2^{(0)}(k)J_1^{(1)}(k) + g_2^{(1)}(k)J_1^{(0)}(k) \right) + \nonumber \\ && \int d^4 p \; F(p) \left[ \bar{J}_1^{(0)}(p)\left( J_2(p,k) + J_2(k,p)\right) +  J_1^{(0)}(p) \tilde{J}_2(k,p) + J_3(p,k,p) + J_3(k,p,p) \right]  , \nonumber \\
&&\tilde{J}_2(k^1,k^2) = 2 \sum_{i = 1}^2 F(k^i) g_2^{(0)}(k^i) \tilde{J}_2(k^1,k^2) + \int d^4 p \; \left( F(p) J_1^{(0)}(p)\left(\bar{J}_3(k^1,p,k^2)+ \bar{J}_3(p,k^1,k^2) \right) \right. \left.+ c.c. \right),\nonumber \\
&&J_2(k^1,k^2) = \sum_{i = 1}^2 F(k^i) g_2^{(0)}(k^i)J_2(k^1,k^2) + \int d^4 p \; F(p) J_1^{(0)}(p) J_3(k^1,k^2,p) , \nonumber\\
&&J_3(k^1,k^2,k^3) = \sum_{i = 1}^3 F(k^i) g_2^{(0)}(k^i)J_3(k^1,k^2,k^3) + 2 \int d^4 p \;  J_1^{(0)}(p) g_4(k^1,k^2,p,k^3) \delta(k^1-k^2+p-k^3).
\end{eqnarray}
\end{widetext}
%%%%%%%%%%%%%%%%%%%%%%%%%%%%%%%%%%%%%%%%%%%%%%%%%%%%%%%%%%%%%%%%%%%%%%%%%%%

%%%%%%%%%%%%%%%%%%%%%%%%%%%%%%%%%%%%%%%%%%%%%%%%%%%%%%%%%%%%%%%%%%%%%%%%%%%%
%%%%%%%%%%%%%%%%%%%%%%%%%%%%%%%%%%%%%%%%%%%%%%%%%%%%%%%%%%%%%%%%%%%%%%%%%%%%

\begin{thebibliography}{99}
\bibitem{seib} N. Seiberg, ``Emergent Spacetime",  arXiv:hep-th/0601234.
\bibitem{Snyder}H.S. Snyder, Phys. Rev. {\bf 71}, 38 (1947).
\bibitem{wit} N. Seiberg and E. Witten, JHEP {\bf 9909}, 032 (1999). 
\bibitem{scholtz} F. G. Scholtz, L. Gouba, A. Hafver, C. M. Rohwer J. Phys. A {\bf 42},175303 (2009).
\bibitem{dop} S. Doplicher, K. Fredenhagen and J. E. Roberts, Commun. Math. Phys. {\bf 172}, 187 (1995).
\bibitem{doug} M. R. Douglas and N. A. Nekrasov, Rev. Mod. Phys., \textbf{73}, 977 (2001).
\bibitem{car1} J. M. Carmona, J. L. Cort{\' e}s, J. Gamboa, and F. M{\' e}ndez, J. High Energy Phys. {\bf 03}, 058 (2003).
\bibitem{car2} J. M. Carmona, J. L. Cort{\' e}s, J. Gamboa, and F. M{\'e}ndez,Phys. Lett. B {\bf 565}, 222 (2003).
\bibitem{pres} V. G\'alikov\'a and P. Pre\v{s}najder, J. Math. Phys. {\bf 54}, 052102 (2013).
\bibitem{scholtz1} N. Chandra, H.W. Groenewald, J.N. Kriel, S. Vaidya and F.G. Scholtz, J.Phys. A \textbf{47}, 445203 (2014).
\bibitem{scholtz2} F. G. Scholtz, J. N. Kriel and H. W. Groenewald, Phys. Rev. D {\bf 92}, 125013 (2015).
\bibitem{scholtz3}  J.N. Kriel, H.W. Groenewald and F.G. Scholtz, Phys. Rev. D. {\bf 95}, 025003 (2017).
\bibitem{bat} M. V. Battisti and S. Meljanac, Phys. Rev. D {\bf  82}, 024028 (2010).
\bibitem{chaichian} M. Chaichian, P. Presnajder and A. Tureanu, Phys. Rev. Lett. \textbf{94}, 151602 (2005).
\bibitem{pinzul} A. Pinzul, J.Phys.A \textbf{45}, 075401 (2012).
\bibitem{car3} J. M. Carmona, J. L. Cort{\'e}s, J. Indur{\'a}in, and D. Maz{\'o}nx,  Phys. Rev. D. {\bf 80}, 105014 (2009).
\bibitem{scholtz4} S. Gangopadhyay and F. G. Scholtz, Phys. Rev. D. {\bf 90},  047702 (2014).
\bibitem{scholtz5} F. G. Scholtz, B. Chakraborty, S. Gangopadhyay, J. Govaerts, J. Phys. A {\bf 38}, 9849 (2005).
\bibitem{scholtz6}S. Gangopadhyay, F. G.Scholtz, Phys. Rev. Lett. {\bf 102} , 241602 (2009).
\bibitem{scholtz7}S. Gangopadhyay, F. G.Scholtz, J. Phys. A: Math. Theor. 47, 075301 (2014).
\bibitem{jackiw} R. Jackiw, 2nd Workshop on Constraint Theory and Quantization Methods, Montepulciano, Italy, June 1993, p.163.
\bibitem{scholtz8} F.G. Scholtz and S. Gangopadhyay, Phys. Rev. D {\bf 71}, 085005 (2005). 
\bibitem{pol}J.~Polchinski, Nucl. Phys. B 231 (1984) 269.
\bibitem{banks}T.~Banks, ``Modern Quantum Field Theory, Cambridge Univ. Press. 2008". 
\bibitem{ros}O.J.~ Rosten, Phys. Rep. 511 (2012) 177.
\end{thebibliography}
\end{document}